\documentclass[useAMS,usenatbib,usegraphicx]{mn2e}

\usepackage{amssymb}
\usepackage{subfigure}
\usepackage{txfonts}

\newcommand{\ie}{\textit{i.e.}}

\newcommand{\ea}{\textit{et al.}}

\newcommand{\eg}{\textit{e.g.}}

\newcommand{\hi}{H{\sc I}}

\newcommand{\nhi}{\mathrm{N_{H I}}}

\newcommand{\lya}{Ly$\alpha$}

\newcommand{\VHf}{Voigt-Hjerting function}

\newcommand{\diff}{\, \mathrm{d}}

	\title[An Analytic Approximation to The Voigt-Hjerting Function ]{Voigt
	Profile	Fitting to Quasar Absorption Lines: An Analytic Approximation to the
	Voigt-Hjerting Function}

	\author[T. Tepper Garc\'\i a]{Thorsten Tepper Garc\'\i a\thanks{E-mail:
		tepper@astro.physik.uni-goe.de
   		}\\
		Institut f\"ur Astrophysik, Georg-August Universit\"at,
		Friedrich-Hund-Platz 1, D-37077 G\"ottingen\\
	}

\begin{document}

	\date{Accepted ---. Received ---; in original form ---.}
	
	\pagerange{\pageref{firstpage}--\pageref{lastpage}}
	
	\pubyear{2005}

	\maketitle
	
	\label{firstpage}
	
	\begin{abstract}
		The \VHf{} is fundamental in order to correctly model the profiles of
		absorption lines imprinted in the spectra of bright background sources
		by intervening absorbing systems. In this work we present a simple
		analytic approximation to this function in the context of absorption line
		profiles of intergalactic \hi{} absorbers. Using basic calculus tools,
		we derive an analytic expression for the \VHf{} that contains only fourth
		order polynomial and Gaussian functions. In connection with the
		absorption coefficient of intergalactic neutral hydrogen, this 
		approximation is suitable for modeling Voigt profiles with an accuracy of
		$10^{-4}$ or better for an arbitrary wavelength baseline, for column
		densities up to $\nhi = 10^{\, 22} \, \mathrm{cm}^{-2}$, and for damping
		parameters $a \lesssim 10^{-4}$, \ie{} the entire range of parameters
		characteristic to all Lyman transitions arising in a variety of \hi{}
		absorbing systems such as \lya{} Forest clouds, Lyman Limit systems and
		Damped \lya{} systems. We hence present an approximation to the \VHf{}
		that is both accurate and flexible to implement in various types of
		programming languages and machines, and with which Voigt profiles can be
		calculated in a reliable and very simple manner.
	\end{abstract}
	


\section{Introduction} \label{sec:intro}
	
	Absorption processes and their signatures (absorption lines) imprinted on
	the spectra of bright background sources (quasars, Gamma-ray bursts, etc.)
	are one of the main sources of information about the physical and chemical
	properties of intervening systems. It is well known that information about
	their temperature, density, chemical abundances, and kinematics can be
	extracted from the analysis of these absorption lines. For instance, a
	detailed insight into the physical state of the intergalactic medium (IGM)
	is provided by the analysis of the absorption lines found in the spectra of
	distant quasars (QSOs) \citep[see \eg{}][]{hue,kim97,kim01,kim02}. These
	lines are due mainly to absorption by neutral hydrogen (\hi{}) present in a
	class of low column density absorbers generally known as \lya{} Forest, and
	due to other elements in low ionisation stages (C{\sc II}, C{\sc IV}, Si{\sc
	II}, Mg{\sc II}, Fe{\sc II}, O{\sc II}, etc.), which arise in higher column
	densities absorbing systems associated with galaxies, such as the Lyman Limit
	Systems (LLSs) and Damped \lya{} Absorbers (DLAs). A wealth of information
	about the distribution, density, temperature, metal content, etc. of these
	systems is now available as a result of exhaustive and extensive studies of
	QSO absorption lines \citep[see \eg{}][ for a review on \lya{} absorbers,
	and on metal systems and DLAs, respectively]{rau,rao}.
	
	In this type of analysis, and within the realm of a given cosmological
	model, the number and observed central wavelength of the absorption lines
	provide information on the spatial distribution of the absorbing systems.
	Furthermore, knowledge about the actual physical state of these systems can
	be obtained basically from the line profiles. Both line counting and line
	profile measurement are tricky tasks though, since the accuracy with which
	they can be performed highly depends on the resolution of the observed
	spectra. For instance, depending on the spatial distribution of the absorbing
	systems, lines can appear very close to each other or even superpose (line
	blending), and a low spectral resolution may lead to the misidentification of
	the resulting composite profile as being a single, complex one. Because of
	this same reason, the determination of the exact shape of each individual
	absorption profile is far from being trivial, and misidentified profiles may
	lead to wrong conclusions about the properties of the absorbing systems.
	
	If one assumes that the physical state of the absorbing medium is uniquely
	defined by its temperature and column density, single absorption line
	profiles are ideally described by Voigt profiles. Mathematically, a Voigt
	profile is given in terms of the convolution of a Gaussian and a Lorentzian
	distribution function, known as \VHf{} \citep{hje}, and a constant factor
	that contains information about the relevant physical properties of the
	absorbing medium (cf. Sect. \ref{sec:abc}). Any departure from a pure
	Voigt profile in the observed lines is expected to yield information about
	the kinematic properties (non-thermal broadening, rotational or turbulent
	macroscopic motions), as well as spatial information (clustering) of the
	absorbing systems.
			
	The \VHf{} has long been known and, consequently, various numerical methods
	to estimate and tabulate this function have been developed and presented
	\citep{hje,har,fin}. Also, a great effort has been done in order to derive
	a semi-analytic approximation to this function that reproduces its
	behavior with high accuracy \citep[see \eg][]{whi,kie,mon}, being the latter
	by far the one with the highest accuracy. With the help of these methods,
	computational subroutines have been developed that make it nowadays possible
	to numerically integrate this function for a wide parameter space \citep[see 
	\eg][]{hum}.
	
	The aim of this work is to make a further contribution to the practical
	handling of the \VHf{} in order to compute Voigt profiles. Starting with an
	exact expression for this function in terms of Harris' infinite series, we
	argue why this series may be truncated to first order in $a$ in the context
	of intergalactic \hi{} absorption lines. We then show that the second term of
	this series can be approximated with a non-algebraic polynomial function,
	which is mathematically simple to handle in the sense that is does not
	contain singularities. Such an analytic, 'well-behaved' expression in terms
	of simple functions as presented here is very attractive, since it allows one
	to replace the many steps and operations needed for numerical integration, or
	to read from look-up tables of values, by a single line with simple
	operations. It is also extremely flexible to implement in various types of
	codes and machines, and is particularly useful for computational routines in
	higher-level programming languages (\eg{} IDL, \emph{Mathematica},
	\emph{Maple}, etc.), in which numerical integration or look-up table reading
	is cumbersome, especially if absorption line profiles have to be calculated
	many times with moderate precision and relative high speed. For instance,
	such an analytic expression should be very useful to synthesise \lya{}
	absorption spectra as in \eg{} \citet{zha,ric}, or in line-fitting algorithms
	like AUTOVP \citep{dav} or FITLYMAN \citep{fon}, used to obtain line
	parameters such as redshift, column density, and Doppler width from
	absorption Voigt profiles imprinted on observed spectra.
	
	In the next section we briefly outline the origin of the Voigt-Hjerting
	function in the Physics of absorption processes, and define the context in
	which the desired approximation of this function is of interest to us. In
	Sections \ref{sec:dawson} and \ref{sec:analyt} we derive this approximation,
	and in Section \ref{sec:ana} we compare the accuracy and speed of a numerical
	method for computing Voigt profiles based on our approximation to other
	existing methods. In Section \ref{sec:app} we present an application of our
	method to model Voigt profiles, and in Section \ref{sec:summ} we summarise
	our main results.

\section{The Voigt-Hjerting Function in the context of \hi{} Absorption Lines}
\label{sec:vhf}
	\subsection{The Absorption Coefficient} \label{sec:abc}

	The probability of a photon with an	energy $E =	h \, c \slash \lambda$ to be
	absorbed within a gas with column density N and kinetic temperature $T$, also
	known as absorption coefficient, is given by
	\begin{equation} \label{eq:tau}
		\tau_{i}(\lambda) = \left(C_i \cdot \mathrm{N} \cdot a\right) \cdot
		H(a,x(\lambda)) \, ,
	\end{equation}
	where
	\begin{displaymath}
		C_i \equiv \frac{4\sqrt{\pi^3}\,e^2}{m_{e} c} \frac{f_i}{\Gamma_i}
	\end{displaymath}
	is a constant for the $i$th	electronic transition caused by the photon
	absorption.	Here $m_{e}$ is the electron mass, $f_{i}$ is the oscillator
	strength, and $\Gamma_i$ the damping constant or reciprocal of the mean
	lifetime of the transition. The function $H$ is the so-called
	\textbf{\VHf{}} and is given by
	\begin{equation} \label{eq:vhf}
		H(a,x) \equiv \frac{a}{\pi} \int_{-\infty}^{+\infty} \,
		\frac{\mathrm{e}^{-y^2}}{(x-y)^2 + a^2} \diff y \, .
	\end{equation}
	Let $\lambda_{i} = h \, c \slash E_i$ be the resonant wavelength of the
	corresponding transition , and $\Delta \lambda_D \equiv \frac{b}{c}
	\lambda_i$ the thermal or Doppler broadening, which defines a Doppler
	unit. Here, the Doppler parameter $b$ is related to the kinetic temperature
	of the gas via $b = \sqrt{2kT / m_p}$, where $k$ is the Boltzmann
	constant and $m_p$ is the proton mass. It follows from these definitions that
	the damping parameter
	\begin{displaymath}
	a \equiv \frac{\lambda_i^2 \Gamma_i}{ 4 \pi c \Delta \lambda_D}
	\end{displaymath}
	quantifies the relative strength of damping broadening to thermal broadening,
	and that the variables $x \equiv \frac{(\lambda - \lambda_i)}{\Delta
	\lambda_D}$ and $y \equiv \frac{\mathrm{v}}{b}$ are just the wavelength
	difference relative to the resonant wavelength in Doppler units, and the
	particle velocity in units of the Doppler parameter, respectively.

	The particular form of the absorption coefficient (\ref{eq:tau}) induces a
	characteristic absorption feature known as Voigt profile. Hence, the Voigt
	profile, and consequently the \VHf{}, naturally arise in the process of 
	absorption-line formation, when one assumes that the physical state of the
	absorbing medium is uniquely defined by its density and kinetic temperature.
	Generally speaking, it is the physical conditions what determines the shape
	of the absorbing features, \ie{} the line profiles. Conversely, it is true
	that line profiles give information about the physical state of the absorbing
	medium. In particular, line profiles other than Voigt profiles give
	information about the departure of the physical conditions assumed here.
		
\begin{figure}
   		\centering
   			\includegraphics[angle=-90, scale=0.35]{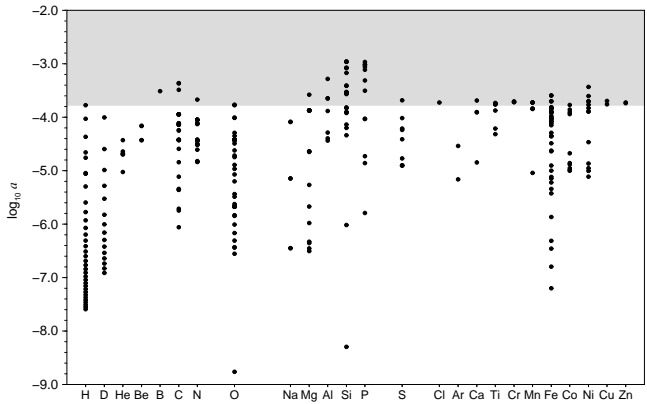}
			\caption[]{\small Value of the damping parameter $a$, assuming
				$b_{\mathrm{HI}} = 36 \, \mathrm{km \, s^{-1}}$, for different
				elements in different ionisation stages typically found in QSO
				spectra. The doppler parameter for other elements is assumed to
				be given by $b_{X} = b_{\mathrm{HI}} \cdot
				\sqrt{m_{\mathrm{HI}}/m_{X}}$ , where $m_{X}$ is the mass of
				element $X$. Elements are listed by increasing atomic mass on
				the $x$-axis, and the logarithmic value of $a$ is given on the
				$y$-axis. For clarity, different ionisation stages of a given
				element have been grouped under a single label. The shaded area
				marks the range above the largest value of $a$ for the
				intergalactic \hi{} Lyman transitions. Note that for the sake of 
				completeness, other elements than those found to date in QSO
				spectra have been included as well.}
			\label{fig:adis}
	\end{figure}

	The class of \hi{} absorbers present in the IGM (\lya{} Forest Clouds) and
	associated with galaxies and larger structures (LLSs, DLAs) can be
	characterised by their column density $\nhi$ and kinetic temperature, and 
	consequently their absorption features observed on \eg{} QSO spectra are well
	described by Voigt profiles. Their observed column densities span a range of
	ten orders of magnitude, approximately from $10^{12} - 10^{22} \,
	\mathrm{cm}^{-2}$, and have temperatures that correspond to Doppler
	parameters approximately in the range $10 - 100 \, \mathrm{km \, s^{-1}}$,
	with a median value around $b_m = 36 \, \mathrm{km \, s^{-1}}$ that decreases
	with redshift \citep{kim97}. For such a range in $b$, the damping parameter
	$a$ for the Lyman transitions of intergalactic \hi{} spans a range of $9.3
	\cdot 10^{-9} - 6.05 \cdot 10^{-4}$. In this case, high values of $a$
	correspond to \lya{}, while lower values are typical for higher order Lyman
	transitions. Transitions of other elements, such as C, Si, Mg, Fe, O, etc.,
	and their various ionisation stages, are also typically found on QSO spectra,
	and their damping parameters cover a range which strongly overlaps with that
	of \hi{}. This can be seen in Figure \ref{fig:adis}, where we show the
	distribution of $a$ for different elements (including \hi{}) in different
	ionisation stages, for a Doppler parameter $b_{\mathrm{HI}} = 36 \,
	\mathrm{km \, s^{-1}}$ for \hi{}, and assuming that the doppler parameter for
	other elements is related to $b_{HI}$ via $b_{X} = b_{\mathrm{HI}} \cdot
	\sqrt{m_{\mathrm{HI}}/m_{X}}$, where $m_{X}$ is the mass of element $X$. For
	clarity, we have grouped all different ionisation stages of a given element
	under a single label. The values of the atomic constants (central wavelength
	of the transition and $\Gamma$-value) have been taken from
	\citet{mor}\footnote{A list containing the values of the damping parameters
	for the elements and their different ionisation stages shown in Figure
	\ref{fig:adis}, is available in plain-text format at
	www.astro.physik.uni-goettingen.de/\~{}tepper/hjerting/damping.dat. Please
	consult this list in order to know the exact value of $a$ for a given element
	in a given ionisation stage.}. The shaded area contains the values of $a$
	above the range characteristic to intergalactic \hi{} for which our
	approximation to the \VHf{}--derived in the next sections--cannot be
	applied or should be applied with caution. Note, however, that the region
	spanned by $a$ for intergalactic \hi{}, \ie{} the region underneath the
	shaded area, contains most of the $a$-ranges spanned by all other elements,
	especially those corresponding to C, O, Mg, and Fe. For other elements, such
	as Cr or Zn, the damping parameter has values right at the upper limit of
	this range.
	For this reason and for the sake of simplicity, in the following we will
	constrain our discussion to the Lyman absorption lines of intergalactic
	\hi{}, but the reader shall bear in mind that the the discussion and method
	to synthesise Voigt profiles presented in this work can be directly applied
	to the transitions of other elements associated with \hi{} absorbing systems,
	according to Figure \ref{fig:adis}.


\subsection{The Absorption Coefficient of \hi{} at Low Column Densities}
\label{sec:low}

	Following \citet{har}, it is true that for $a < 1$, \ie{} when Doppler
 	broadening dominates over damping broadening, the Voigt-Hjerting
	function (\ref{eq:vhf}) can be expressed as
	\begin{equation} \label{eq:vhf2}
		H(a,x) = \sum_{n=0}^{\infty} H_n\,(x) \, a^n \, ,
	\end{equation}
	where the functions $H_n(x)$ are defined by
	\begin{equation} \label{eq:vhf3}
		H_n\,(x) \equiv \frac{(-1)^n}{\sqrt{\pi}\,n!} \int_{0}^{\infty} v^n
		\mathrm{e}^{-(v\slash2)^2} \cos(xv) \diff v \, .
	\end{equation}
	These functions are bounded with respect to $n$ and $x$, and they have
 	values of the order of unity. Indeed, taking the absolute value of the 
	integral, neglecting the cosine, which takes values of the order of unity,
	and computing the resulting integral one can show that
	\begin{equation} \label{eq:h_n}
		\vert H_n(x) \vert \leq \frac{2}{\sqrt{\pi}} \approx 1.123 \, ,
	\end{equation}
	for all $n \in \{0,1,2, \ldots\}$ and $x \in \mathbb{R}$. From this it
	follows that if $a \ll 1$, the \VHf{} can very well be approximated to
	zeroth order in $a$ by the first term of the series (\ref{eq:vhf2}), \ie{}
	\begin{math}
		H(a,x) \approx H_0(x) \, ,
	\end{math}
	as first noted by \citet{wal}. Note that this result is \emph{exact} in
	the limit $a \to 0$. Taking the definition (\ref{eq:vhf3}), it follows that
	\begin{equation} \label{eq:h_0}
		H_0\,(x)\ = \,\mathrm{e}^{-x^2} \, ,
	\end{equation}
	and thus
	\begin{math}
		H(a,x) \approx \mathrm{e}^{-x^2},
	\end{math}
	for $a \ll 1$. We call this the \VHf{} to \emph{zeroth order}.
	
\begin{figure}
   		\centering
   			\includegraphics[angle=-90, scale=0.35]{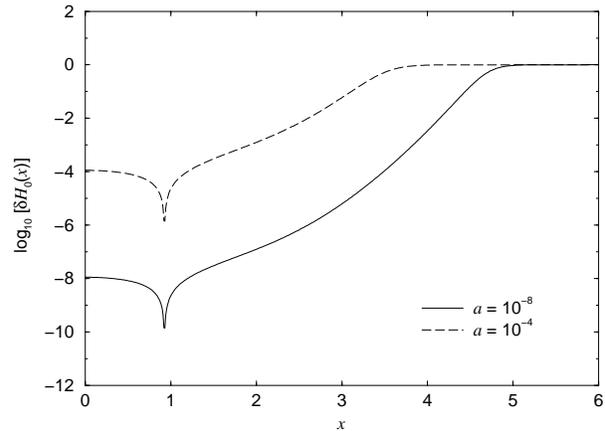}
			\caption[Comparison of $H_0$ and $H$.]{\small Departure of the
				\VHf{} from a pure Gaussian function for $a = 10^{-8}$ (solid
				line) and $a = 10^{-4}$ (dashed line) as a function of $x$. The
				departure is given as the logarithmic difference between the
				\VHf{} and the zeroth order approximation in the form
				\mbox{$\delta H_0 \equiv 1 - H_0\slash H$}. The greater this
				quantity, the less accurate is the zeroth order approximation.}
			\label{fig:HvsH0}
	\end{figure}

	But what actually means that the condition $a \ll 1$ be satisfied, so that
	this zeroth order approximation can be safely used to model absorption line
	profiles? In order to address this, we compute for the extreme values of $a$
	for intergalactic \hi{}, $a \approx 10^{-8}$ and $a \approx 10^{-4}$, the
	departure of the \VHf{} from a pure Gaussian function. This is shown in
	Figure \ref{fig:HvsH0} as the logarithmic difference between\footnote{The
	values of the \VHf{} were computed numerically using a routine based on
	Monaghan's algorithm in \citet{mur} (cf. Section \ref{sec:ana}).} $H$ and
	$H_0$ relative to $H$ as a function of $x$, \ie{} the quantity \mbox{$\delta
	H_0 \equiv 1 - H_0/H$}. Note how the zeroth order approximation completelly
	differs from the actual \VHf{} at $x \gtrsim 3.5$ for $a = 10^{-4}$, and at
	$x \gtrsim 4.5$ for $a =10^{-8}$. Thus, even a value of $a$ as small as
	$10^{-8}$ is not a sufficient condition for the zeroth order term to be a
	good approximation of $H$ for an arbitrary range in $x$.

	In the context of the \hi{} absorption coefficient, the condition on $a$ for
	which the approximation to zeroth order of the \VHf{} is valid, translates
	into a restriction on $\nhi$, and more specifically on the quantity $C_i
	\cdot a \cdot \nhi$. This can be seen by taking a glance at equation
	(\ref{eq:tau}): Despite the fact that the condition $a \ll 1$ holds, the
	product $C_i \cdot a \cdot \nhi$ can be very large for high
	enough\footnote{'High enough' means in this case that the condition $\nhi >
	(C_i \cdot a)^{-1}$ is satisfied.} column densities, and since the functions
	$H_n$ are of the order of unity, such terms can have a significant
	contribution to the absorption coefficient. Along this line of reasoning,
	and taking into consideration that the constant $C_i$ in equation 
	(\ref{eq:tau}) is of the order of $10^{-11} \, \mathrm{cm}^{2}$ for all
	Lyman transitions, it is clear that the departure of the \VHf{} from its
	zeroth order approximation becomes significant in the wavelength ranges $x
	\gtrsim 3.5$ and $x \gtrsim 4.5$ (in Doppler units) at column densities
	$\nhi \gtrsim 10^{15} \, \mathrm{cm}^{-2}$ for a damping parameter $a =
	10^{-4}$, and at column densities $\nhi \gtrsim 10^{19} \, \mathrm{cm}^{-2}$
	for $a = 10^{-8}$. Since intervening \hi{} absorbers typically have
	column densities in the range $10^{12} - 10^{22} \, \mathrm{cm}^{-2}$,
	a Gaussian approximation to $H$ for modeling intergalactic \hi{} absorption
	line profiles is only suitable for the low end of the column density range.
	
	In addition to the factor $C_i \cdot a \cdot \nhi$ being large and even
	more decisive for the uselessness of the zeroth order approximation for an
	arbitrary range in $x$, is the fact	that the zeroth order term,
	$\mathrm{e}^{-x^2}$, rapidly decreases for large values of $x$ and is
	therefore overwhelmed by higher order terms in Harris' expansion, which hence
	dominate the behavior of the absorption coefficient, as already seen in
	Figure \ref{fig:HvsH0}. To shed some light on this, consider the following
	numerical example: Out to $x \approx 4$ (in Doppler units), and for $a =
	10^{-4}$, the \VHf{} is of the order of $4.2 \cdot 10^{-6}$. The zeroth order
	term in the series (\ref{eq:vhf2}) satisfies $H_0~(x = 4) \approx 1.1 \cdot
	10^{-7}$, whereas the first order term $(a \cdot H_1)~(x = 4) \approx 3.9
	\cdot 10^{-6}$. Thus, at large enough $x$, the behaviour of $\tau$ is
	evidently governed by the terms of order $n \geq 1$ in the series
	(\ref{eq:vhf2}).


\subsection{Higher Column Densities and First Order Term} \label{sec:high}
	
	Due to the arguments stated above, and even though the damping parameter
	satisfies $a \ll 1$, it is clear that the absorption coefficient of
	intergalactic \hi{} cannot simply be approximated by a constant times
	$\mathrm{e}^{-x^2}$ for an arbitrary wavelength range and column densities
	grater than $10^{15} \, \mathrm{cm}^{-2}$. One actually has to take into
	account terms of higher order in the series (\ref{eq:vhf2}), at least to
	first order in $a$, \ie{}
	\begin{math}
		H(a,x) \approx (H_0 + a \cdot H_1)~(x) \, .
	\end{math}
	In fact, one should take into account all terms up to $N$th order for values
	of $(C_i \cdot a \cdot \nhi)^{-1}$ that are nearly equal or greater than the
	absolute difference between the sum
	\begin{math}
		\sum_{n=0}^{N} H_n\,(x) \, a^n \, ,
	\end{math}
	and the exact \VHf{}. However, as we shall show next, the approximation to
	first order in $a$ in Harris' expansion is enough to model absorption line
	profiles with moderate to high accuracy for the range of parameters
	$(a,\,\mathrm{N_{HI}},\,C_i)$ characteristic to intervening \hi{} absorbers.

\begin{figure}
   		\centering
   			\includegraphics[angle=-90, scale=0.35]{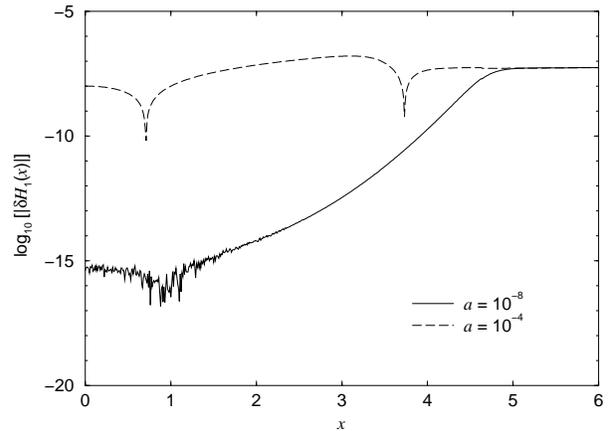}
			\caption[Comparison of $H_0$ and $H_1$.]{\small Contribution from
				the zeroth and first order terms of the series (\ref{eq:vhf2})
				to the absorption coefficient, when $(C_i \cdot a \cdot \nhi)$
				is of the order of unity. The curves show the logarithmic
				difference between $H$ and $H_0 + a \cdot H_1$, relative to $H$
				as a function of $x$ for $a = 10^{-8}$ (solid line) and $a =
				10^{-4}$ (dashed line). The smaller this difference, the greater
				the contribution from the zeroth and first order term to $H$.
				Note that the behavior of $H$ is indeed dominated with a
				difference of seven or more orders of magnitude by the first two
				terms of Harris' expansion for the whole range in $x$ shown here.
			    The values for $H_1$ were calculated according to equation
				(\ref{eq:h_1}), Section \ref{sec:dawson}, using	numerical
				integration to compute the function $F$.}
			\label{fig:HvsH0H1}
	\end{figure}

	In order to prove the above statement, we look at the relative contribution
	of the zeroth and first order terms to the \VHf{} for $a \in \{10^{-8},
	10^{-4}\}$ and an arbitrary range in $x$. This is achieved, for example, by
	computing the logarithm of the quantity $\delta H_1 \equiv \left| 1 - (H_0 +
	a \cdot H_1)/H \right|$ as a function of $x$ for the two extreme values of
	$a$, as shown in Figure \ref{fig:HvsH0H1}. Here we have advanced the function
	$H_1$ which is being handled in the next section. Note that the greater the
	contribution from the zeroth and first order term to $H$, the smaller the
	quantity $\delta H_1$. In this case, as can clearly be seen, $\delta H_1$
	takes on values of the order of $10^{-7}$ or less over the whole wavelength
	range shown here and for the whole range in $a$ for intergalactic \hi{}.
	
	If one takes into account that $H(x) \leq 1$ for all $x$, it is obvious that
	the relative difference is equal or greater than the \emph{absolute}
	difference, \ie{} $\delta H_1 \geq \left| H - (H_0 + a \cdot H_1) \right|$.
	Thus, in the case of $a \approx 10^{-4}$ and $\delta H_1 = 10^{-7}$, the
	departure of $H$ from its first order approximation becomes significant
	at column densities $\nhi > (a \cdot \delta H_1 \cdot C_i)^{-1} = 10^{22} \,
	\mathrm{cm}^{-2}$, and at even larger $\nhi$ for $a < 10^{-8}$ and/or
	smaller $\delta H_1$. Hence, the first two terms of Harris' expansion
	dominate the behaviour of $H$ over the whole wavelength range shown here,
	with an accuracy of $10^{-7}$ or greater, for the range in $a$ characteristic
	to intergalactic \hi{}. On this basis, we consider thar an approximation to
	first order in $a$ of the \VHf{} in terms of the functions $H_0$ and $H_1$ is
	suitable to model Voigt profiles.
	Since the function $H_0$ is known and simple \emph{per se}, we now turn to
	the task of finding an approximation to the function $H_1$ in terms of a
	simple, analytic expression.

\section{The Dawson Function Revisited} \label{sec:dawson}

	According to definition (\ref{eq:vhf3}) we have
	\begin{equation} \nonumber
		H_1(x) = \frac{-4}{\sqrt{\pi}} \int_{0}^{\infty} v \,
		\mathrm{e}^{-v^2} \cos(2xv) \diff v \, .
	\end{equation}
	Integrating this equation partially, and computing the resulting Sinus
	transform of a Gaussian it follows \citep{mih}
	\begin{equation} \label{eq:h_1}
		H_1(x) = \frac{-2}{\sqrt{\pi}} \Big [1 - 2x \, F(x) \Big ] \, ,
	\end{equation}
	where we adopt the notation first introduced by \citet{mil}
	\begin{equation} \label{eq:daw}
		F(x) \equiv \mathrm{e}^{-x^2} \int_{0}^{x} \mathrm{e}^{v^2} \diff v \, .
	\end{equation}
	$F$ is known as the \emph{Dawson function} \citep{daw}.
	
	It is evident that finding an approximation for $H_1$ translates into the
	same problem for $F$. We thus want to show that	a simple analytic expression
	can be found which approximates the Dawson function, and hence the function
	$H_1$, accurately enough, in order for an approximation of the \VHf{} in
	terms of these functions to be useful for synthesising Voigt profiles.
\subsection{Properties of the Dawson Function} \label{sec:prop}

	We briefly want to state some important properties of the Dawson-Function.
	First, this function is antisymmetric, \ie{} $F(-x) = - F(x)$ for all $x \in
	\mathbb{R}$. Because of this, from now on we restrict our analysis to $x
	\geq 0$. Besides, it has no roots in the positive semi-axis, and $F(0) = 0$,
	as can easily be seen from its definition (Fundamental Theorem of Calculus).
	Furthermore, $F$ is bounded, since $H_1$ is bounded itself (cf. Sect.
	\ref{sec:low}). Indeed, differentiation with respect to $x$ gives
	\begin{equation} \label{eq:daw2}
		\frac{\diff}{\diff x} F(x)  = 1 - 2x \, F(x) \, .
	\end{equation}
	Hence, the upper bound is given by $F(x_0) = (2x_0)^{-1}$, where $x_0$ is
	defined by equating $F$'s derivative to zero and solving for $x$. Actually,
	$F$ has its maximum at $x_0 = 0.92413$ with $F(0.92413) = 0.54104$ \citep[see
	\eg{}][]{hmf}.	From this last equation it follows also that equation
	(\ref{eq:h_1}) can be rewritten as
	\begin{equation} \label{eq:h_12}
		H_1(x) = \frac{-2}{\sqrt{\pi}} \frac{\diff}{\diff x} F(x) \, .
	\end{equation}

	We want to know how the function $F$ behaves asymptotically, \ie{} for $x
	\ll 1$ as well as for $x \gg 1$. Using the power series of the exponential
	we get
	\begin{equation} \label{eq:daw3}
		F(x) = \mathrm{e}^{-x^2} \int_{0}^{x} \sum_{n=0}^{\infty} \frac{1}{n!}
		v^{2n} \diff v = \mathrm{e}^{-x^2}\cdot\sum_{n=0}^{\infty} \frac{1}{n!}
		\frac{x^{2n+1}}{2n+1} \, .
	\end{equation}
	In this case, the sum and the integral operator commute, since the power
	series of the exponential converges uniformly in any interval $[a,b]$,
	particularly for $v \in [0,x]$ \citep[see \eg{}][]{for}. Expressing the term
	$\mathrm{e}^{-x^2}$ by its corresponding power series and rearranging terms,
	this last equation reads in explicit form
	\begin{equation} \nonumber
		F(x)  = x\cdot\Big(1-\frac{2}{3}x^2+\frac{4}{15}x^4+\cdots \Big) \, .
	\end{equation}
	Thus, for $x \ll 1$ the Dawson function behaves asymptotically up to third
	order as
	\begin{equation} \label{eq:fap1}
		F(x) \approx x\cdot(1-\frac{2}{3} x^2), \qquad x \ll 1 \, .
	\end{equation}
	In order to investigate how $F$ behaves for $x \gg 1$, we first rewrite the
	function (\ref{eq:daw}) with the replacement $v'=x-v$ and the aid of the
	power series of the exponential as
	\begin{equation} \label{eq:daw4}
		F(x) = \sum_{n=0}^{\infty} \frac{1}{n!} I_n(x) \, ,
	\end{equation}
	with the definition
	\begin{equation} \label{eq:I_n}
		I_n(x) \equiv \int_{0}^{x} v^{2n} \mathrm{e}^{-2xv} \diff v \, .
	\end{equation}
	It is not hard to see that every term of the series (\ref{eq:daw4}) is 
	separately bounded with respect to $x$ as well as $n$. Making the
	replacement $v'=2vx$ in eq. (\ref{eq:I_n}), and integrating for $x \neq 0$
	we get for $n \in \mathbb{N}_0$
	\begin{equation} \label{eq:I_n2}
		I_n(x) = \frac{1}{(2x)^{2n+1}}(2n)! (1 - \mathrm{e}^{-2x^2}) -
		\mathrm{e}^{-2x^2} R_n(x) \, ,
	\end{equation}
	where $R_n(x)$ is a rather cumbersome polynomial function. Now, for $x \gg
	1$, we may drop all terms which contain an exponential factor and in this way
	we get the asymptotic form
	\begin{equation} \label{eq:fap2}
		F(x) \approx \sum_{n=0}^{\infty} \frac{1}{(2x)^{2n+1}}(2n)!, \qquad x
		\gg 1 \, .
	\end{equation}
	It can be seen from this expression that $F$ vanishes as $(2x)^{-1}$ for $x
	\to \infty$, and that the first derivative (\ref{eq:daw2}), and thus the
	function $H_1$, also vanish in this limit as $(2x^2)^{-1}$. From equation 
	(\ref{eq:daw2}) it is also true that $F'$ converges to unity for $x \to 0$
	and that $H_1$ converges to $(-2 \ / \sqrt{\pi})$ in this limit. Since we
	want our approximation to $F$, and consequently to $H_1$ and $H$, to be valid
	in the whole range $x \in [0,\infty)$, we require it to fulfill both these
	conditions as well.


\section{The analytic approximation $D_1$} \label{sec:analyt}

	Equation (\ref{eq:daw4}), together with eq. (\ref{eq:I_n}), represent indeed
	an exact expression for the Dawson function. However, these expression are
	not suitable for practical computation. We therefore explore the possibility
	of finding an analytic expression which is easy to handle and which can	be
	used to compute the value of $F(x)$	for $x \in [0,\infty)$. In particular, we
	shall see if it is possible to truncate the series (\ref{eq:daw4}) in order
	to find an approximation to $F$, which has all its properties (antisymmetry,
	boundedness, etc.), which converges for $x \to \infty$ as well as for $x \to
	0$, and which is well defined in the whole range $[0,\infty)$. For instance,
	equations (\ref{eq:fap1}) and (\ref{eq:fap2}) do not fulfill these
	requirements. Nevertheless, they show us how our desired function has to
	behave asymptotically.
	
	Let us define
	\begin{equation} \label{eq:fap3}
		D_N(x) \equiv \sum_{n=0}^{N} \frac{1}{n!} I_n(x) \, .
	\end{equation}
	where the $I_n$'s are given by eq. (\ref{eq:I_n}). Using this definition
	we get
	\begin{equation} \label{eq:d1}
		D_1(x) = (1 - \mathrm{e}^{-2x^2}) \cdot \left[\frac{1}{2x} +
		\frac{1}{4x^3} \right]  - \mathrm{e}^{-2x^2} \cdot \left[\frac{x}{2} +
		\frac{1}{2x} \right] \, .
	\end{equation}
	It is easy to show that this function behaves qualitatively in the same way
	as $F$ does, \ie{} it is antisymmetric, bounded, and has no roots in the
	positive semi-axis. Furthermore, both these functions have the same
	asymptotical behavior. Indeed, up to third order we have for $x \ll 1$
	\begin{equation} \label{eq:d1_2}
		D_1(x) \approx x \cdot (1 - \frac{2}{3} x^2), \qquad x \ll 1 \, .
	\end{equation}
	as can be shown by expanding the exponentials in eq. (\ref{eq:d1}) in
	terms of their power series. A glance at eq. (\ref{eq:fap1}) makes the
	similarity between $F$ and $D_1$ evident in this limit. For $x \gg 1$ we get
	from eq. (\ref{eq:d1}), neglecting the exponentials,
	\begin{displaymath}
		D_1(x) \approx \frac{1}{2x} + \frac{1}{2^2x^3}, \qquad x \gg 1 \, ,
	\end{displaymath}
	that is the same as for $F$ (see eq. \ref{eq:fap2}). Furthermore, the first
	derivative of $D_1$ with respect to $x$ is unity at $x = 0$ and vanishes as
	$(2x^2)^{-1}$ for $x \to \infty$. However, the maximum of $D_1$ is at $x =
	0.87269$ with $D_1(0.87269) = 0.52212$, \ie{} at slightly different values
	from those of $F$.
	
	The magnitude and range of the error in approximating the Dawson function by
	the function $D_1$ can be estimated, at least qualitatively, in the following
	way: From equations (\ref{eq:daw4}) and	(\ref{eq:fap3}) it follows that
	\begin{equation} \label{eq:fap4}
		F(x) = D_1(x) + \sum_{n=2}^{\infty} \frac{1}{n!} I_n(x) \, .
	\end{equation}
	The sum in this last equation, \ie{} the absolute error in our approximation
	$F(x) \approx D_1(x)$, is a positive semi-definite quantity, since each term
	of the sum has this property. Hence, it is true that $0 \leq D_1(x) \leq
	F(x)$ for $x \in [0,\infty)$. Furthermore, since both $F$ and $D_1$ converge
	to zero for $x \to 0$ as well as for $x \to \infty$, and both these functions
	are bounded, it follows from eq. (\ref{eq:fap4}), that the error also
	vanishes for $x \to 0$ as well as for $x \to \infty$, that it is also
	bounded, and that its maximum value is less equal than ${max} \left\{ F(x) -
	D_1(x) \right\}_{x > 0}$. From this, and since their respective maxima are
	slightly shifted with respect to each other, one is led to the conclusion,
	that the error is constrained to a narrow range in $x$ and that the maximum
	error in our approximation to $F$ occurs near the maxima of these functions,
	\ie{} in the vicinity of $x = 1$.
	
	We wont further try to quantify the actual error in our approximation to $F$.
	It shall be enough to know, for our purposes of finding an approximation to
	the \VHf{}, that the error in the approximation $F(x) \approx D_1(x)$ is
	bounded and constrained to a narrow wavelength interval around $x = 1$.
	Besides, we will indirectly estimate the error in this  approximation when
	quantifying the error in our approximation to $H$ in Section \ref{sec:ana}.


\subsection{The \VHf{} to First Order} \label{sec:first}

	Once we have found an approximation to the Dawson function, we can use it to
	give the desired expression for the \VHf{} using Harris' expansion to first
	order in $a$. Replacing in equation (\ref{eq:h_12}) the function $F$ by our
	approximation $D_1$ (eq. \ref{eq:d1}), taking the corresponding derivative,
	and rearranging terms we get
	\begin{equation}  \label{eq:h_13}
		H_1(x) \approx \frac{-2}{\sqrt{\pi}} K(x) \, \mathrm{e}^{-x^2} \, ,
	\end{equation}
	where we have defined
	\begin{equation} \label{eq:h_14}
		K(x) = \frac{1}{2 x^2} \left[ (4 x^2 + 3) \, (x^2 + 1)
		\, \mathrm{e}^{-x^2} - \frac{1}{x^2} (2 x^2 + 3) \sinh x^2 \right] \, .
	\end{equation}
	We want to highlight the fact that equation (\ref{eq:h_13}) is well defined,
	\ie{} it has no singularities in the whole interval $[0,\infty)$.
	Furthermore, it converges to the correct value in the limits $x \to 0$
	and $x \to \infty$. Indeed, it is easy to show that
	\begin{math}
		\lim_{x \to 0} H_1(x) = \frac{-2}{\sqrt{\pi}} \, ,
	\end{math}
	and
	\begin{math}
		\lim_{x \to \infty} H_1(x) = 0 \, ,
	\end{math}
	whether one uses for $H_1$ the exact expression (\ref{eq:h_1}) or the
	approximation (\ref{eq:h_13}). In contrast, in approximations to the
	\VHf{} to model Voigt profiles often used in the literature
	\citep[see \eg{}][]{spi,zha} and given in the form
	\begin{math}
		c_1 \cdot \mathrm{e}^{-x^2} + c_2 \cdot \frac{1}{x^2} \, ,
	\end{math}
	where the $c_i$'s are constants, the second term which represents the
	Lorentzian damping clearly diverges for $x \to 0$, and one has to
	artificially define the wavelength range in which this second term is used.
	It is customary to neglect this term for low column densities and in the
	vicinity of $x = 0$, but how to exactly choose the radius of the vicinity is
	not clear and completely arbitrary. However, with an expression like equation
	(\ref{eq:h_13}) at hand, no such assumptions have to be made.
	
	Taking into account that $a \ll	1$ in order to neglect terms of order $n \geq
	2$ in the series (\ref{eq:vhf2}), we get, using the expressions for	$H_0$
	(eq. \ref{eq:h_0}) and $H_1$ (eq. \ref{eq:h_13}), that the \VHf{} to first
	order in $a$ is given by
	\begin{equation} \label{eq:vhf4}
		H(a,x) \approx \mathrm{e}^{-x^2} \left[1 - a \frac{2}{\sqrt{\pi}} \, K(x)
		\right] \, .
	\end{equation}
	This expression is symmetric in $x$, as it should be, and thus it is valid
	for $x \in \mathbb{R}$ and $a \ll 1$. According to this equation, the \VHf{}
	can be regarded as a ''corrected'' Gaussian function, where the correction
	term depends on the parameter $a$. In the context of the absorption
	coefficient of \hi{}, this correction term also depends on the column
	density $\nhi$, of course, via the quantity $a \cdot \nhi$.


\section{Analysis} \label{sec:ana}

	In order to quantify the quality of our approximation to $H$, we perform a
	test on speed as well as on precision, comparing a numerical method to
	compute $H$, based on our approximation, to other standard, available methods
	to numerically compute this function. For this purpose, we use the approach
	and corresponding computational routine developed by \citet{mur}, which
	consists of the numerical implementation in FORTRAN of four different methods
	to compute $H$: Harris' $H1$ and $H2$, Huml\'\i$\check{\textnormal{c}}$ek's,
	and Monaghan's. In Murphy's notation, Harris' $H1$ and $H2$ correspond to the
	\VHf{} approximated by the first three and five terms of the series expansion
	(\ref{eq:vhf2}), respectively. Huml\'\i$\check{\textnormal{c}}$ek's optimized
	algorithm and Monaghan's differential approach to approximate $H$ are
	explained in detail in \cite{hum} and \cite{mon}, respectively. Our method to
	compute $H$	consists simply in the numerical implementation in
	FORTRAN\footnote{The rearrangement of equations (\ref{eq:h_14}) and
	(\ref{eq:vhf4}) that leads to the smallest number of operations reads, in
	code syntax,
	\begin{displaymath}
		H(a,x) = H_0  - a \, \slash\sqrt{\pi} \, \slash x^2 \cdot \left[H_0 \cdot
		H_0 \cdot \left(4 \cdot x^2 \cdot x^2 + 7 \cdot x^2 + 4 + Q\right) - Q -
		1 \right] \, ,
	\end{displaymath}
	where the terms $x^2$, $H_0 \equiv \mathrm{e}^{-x^2}$ and $Q \equiv 1.5
	\cdot x^2$ have to be computed just once.}
	of equations (\ref{eq:h_14}) and (\ref{eq:vhf4}).

\subsection{Speed}
	
	Following Murphy's approach, the relative speed of all five methods are
	determined by calculating the time that a routine based on each method
	requires to compute the \VHf{} for \mbox{$x \in [0,10]$} and damping
	parameters $a$ in the range $10^{-8} - 10^{-4}$ for a total of $1.5 \cdot
	10^{7}$ runs. In this way, we get that the relative speed\footnote{This
	calculations were performed on an Intel Xeon 3.2 GHz processor.} of each
	method in the order \mbox{$H1$ : (this work) :
	Huml\'\i$\check{\textnormal{c}}$ek : $H2$ : Monaghan} corresponds to
	\mbox{1 : 4.2 : 5.8 : 6.8 : 66.6}, independent of $a$. According to this
	result, our method is second fastest.

\subsection{Precision}

	\begin{figure*}
		\begin{center}
			\vspace{0.5cm}
			\includegraphics[width=0.9\linewidth,angle=-90,scale=0.85]{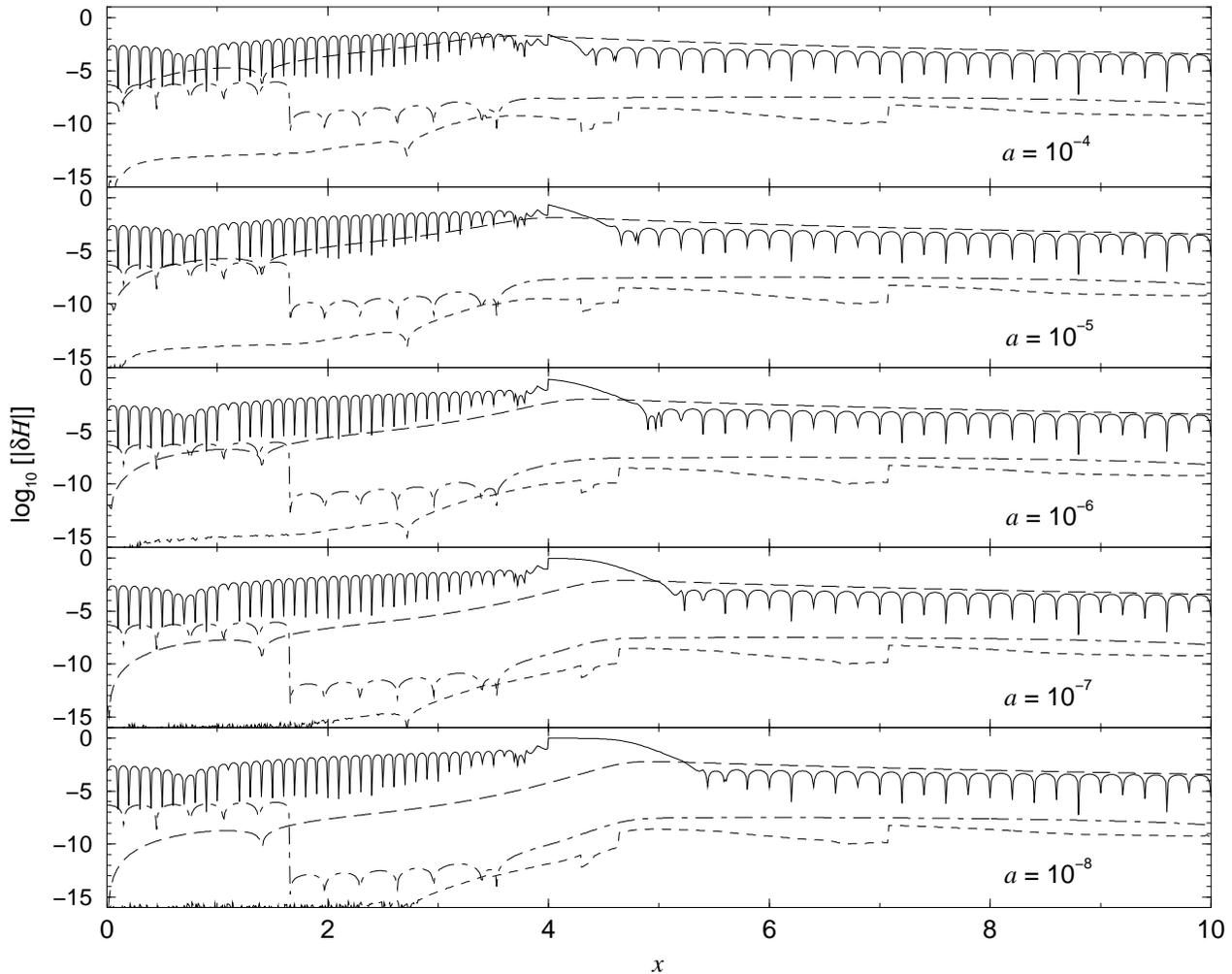}
		\end{center}
		\vspace{0.5cm}
		\caption[]{\small Precision of different methods to compute the \VHf{},
			relative to Monaghan's differential method. Shown here is the
			logarithmic difference as a function of $x$, \ie{} the quantity
			$\log_{10} \, \delta H$, with $\delta H \equiv 1 -
			H_{method}/H_{Mon}$, where $H_{Mon}$ and $H_{method}$ are,
			respectively, the \VHf{} computed using Monaghan's algorithm and each
			of the methods mentioned in the text: H1 (solid line), H2
			(short-dashed line), Huml\'\i$\check{\textnormal{c}}$ek (dot-dashed
			line), this work (long-dashed line). Each panel corresponds to a
			different damping parameter. Here we chose the range of $a$
			characteristic to intergalactic \hi{}, \ie{} $a \in [10^{-8},
			10^{-4}]$. This graph was created using the approach and
			corresponding routine developed by \citet{mur}, and is adapted from
			Figure A.1 of the same work.}
		\label{fig:comp}
	\end{figure*}
	The precision of our method relative to the other methods mentioned above is
	determined in the following way: First, the values of the \VHf{} computed
	using Monaghan's algorithm for $x \in [0,10]$ and $a \in [10^{-8},10^{-4}]$
	are taken as fiducial. Then, the value of $H$ for the same range in	$x$ and
	$a$ is computed using each of these methods, and the logarithmic difference
	between each method and its fiducial value, \ie{} the quantity $\log_{10} \,
	\delta H$ with $\delta H \equiv 1 - H_{method}/H_{Mon}$, is calculated as a
	function of $x$ for each different $a$. The result is shown in Figure 
	\ref{fig:comp}.	It can be seen from this figure that Harris' $H2$ is the
	second most precise method to compute $H$, if one takes Monaghan's algorithm
	as fiducial, but is six times slower than Harris' $H1$, and one-and-a-half
	times slower than our method, as stated in the previous section. Our method
	has a precision of $10^{-4}$ or better for $x \lesssim 4$. For all values of
	$a$, the difference peaks around $x = 4$ to a value of the order of 0.01, and
	the precision increases again for values of $x > 4$. The precision is better
	for smaller $a$, as expected, since the zeroth order term gains in importance
	in our approximation for decreasing $a$. For $a \lesssim 10^{-6}$ and $x <
	1.5$, our method is more precise than Harris' $H1$ or
	Huml\'\i$\check{\textnormal{c}}$ek's, and, as seen above, 1.5 times	faster
	than the latter.
	
\subsection{Modeling of \hi{} absorption profiles} \label{sec:mod}
	
\begin{figure}
   		\centering
   			\includegraphics[angle=-90, scale=0.35]{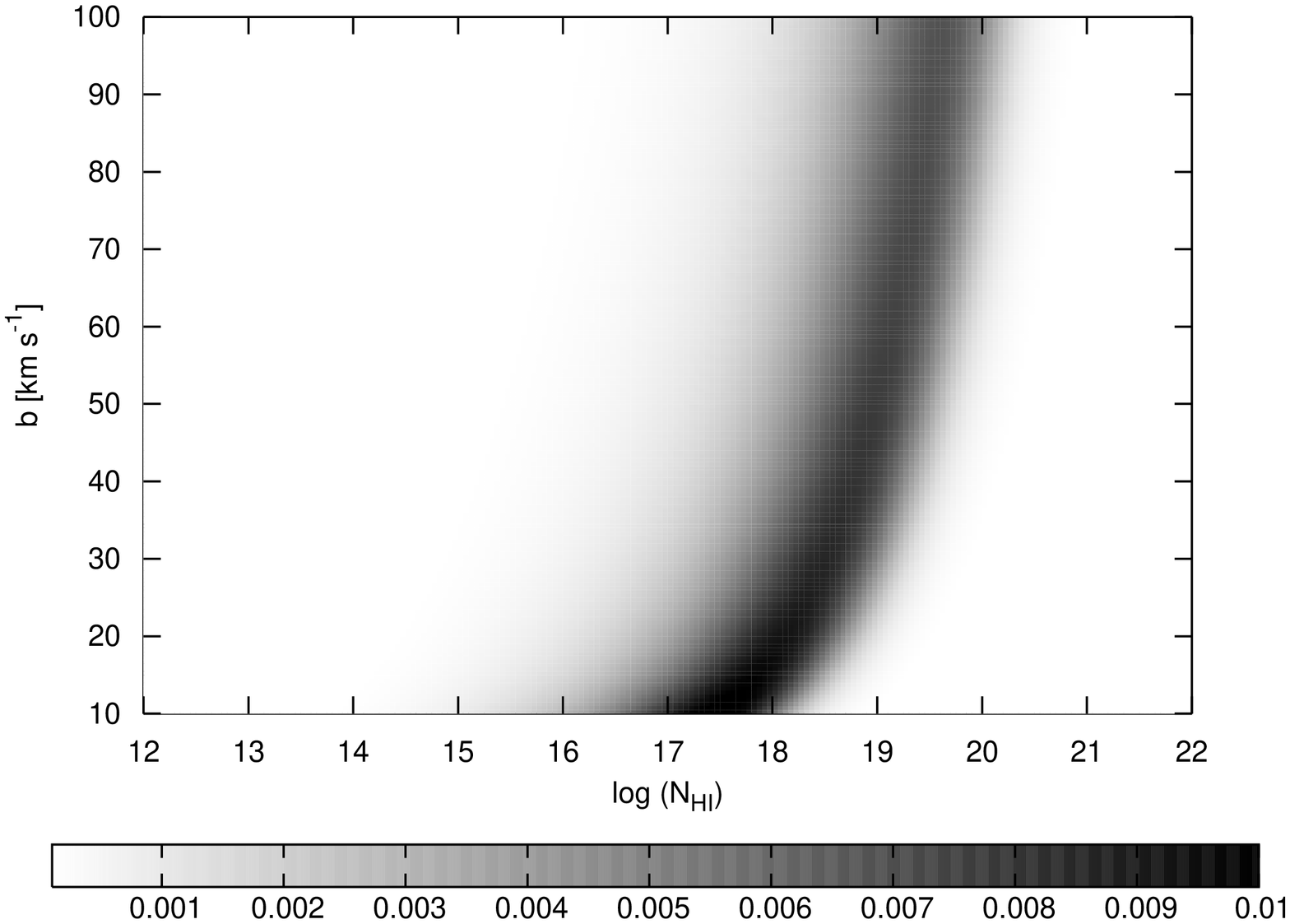}
			\caption[]{\small \emph{Worst Scenario}: Lower limit to the precision
				of our method to synthesise Voigt profiles, given as the maximum
				value of the difference $\delta V $ (cf. text for definition)
				between a \lya{} absorption profile computed according to
				equation (\ref{eq:tau}) using our approximation to the \VHf{}
				(eq. \ref{eq:h_14} and \ref{eq:vhf4}) and Monaghan's algorithm,
				for the whole range of values for the parameters $(\log \nhi,b)$
				characteristic to intergalactic \hi{}. The value corresponding to
				each pair $(\log \nhi,b)$ is the \emph{maximum} value of the
				quantity $\delta V$ in the \emph{entire} wavelength range
				$\lambda \in [100,2300]$ \AA{}. For clarity, any value below
				$10^{-4}$ has been coded as zero.}
				\label{fig:map}
	\end{figure}

	We now turn to analyse how accurate is our method in order to model \hi{}
	absorption profiles. Taking the whole range in column density $\log \nhi \in
	[12.0,22.0]$ dex and Doppler parameters $b \in [10.0,100.0] \, \mathrm{km \,
	s^{-1}}$ characteristic to intergalactic \hi{}, we synthesise for each pair
	$(\nhi,b)$ (with a resolution of $\Delta \log \nhi = 0.05$ dex, and $\Delta b
	= 0.5 \, \mathrm{km \, s^{-1}}$) a single \lya{} absorption profile in the
	range $\lambda \in [100,2300]$ \AA{} with a resolution of $\Delta \lambda  =
	0.01$ \AA{}. The absorption profile is synthesised according to equations
	(\ref{eq:tau}) and (\ref{eq:vhf}), using both our method and Monaghan's to
	compute $H$. We then compute for each pair $(\nhi,b)$ the absolute value of
	the difference between the profiles generated using these two methods
	relative to Monaghan's, \ie{} $\delta V \equiv |1 - \mathrm{e^{-\Delta
	\tau}}|$, with $\Delta \tau \equiv \tau_{our} -\tau_{Mon}$, as a function of
	wavelength for the whole wavelength range, and pick the \emph{maximum} value
	of this difference in the \emph{entire} range. We choose to do so in order to
	pick up the worst cases possible, \ie{} those with the lowest accuracy, and
	put in this way a stringent lower limit to the accuracy of our method. The
	result is shown as a contrast diagram on the $(\log \nhi,b)$-plane in Figure
	\ref{fig:map}. Note that, in this case, it is not the logarithmic, but the
	\emph{linear} difference which is shown. The highest precision is of the
	order of $10^{-16}$ or even better. However, for the sake of simplicity, any
	value below $10^{-4}$ has been coded as zero. The largest discrepancy between
	both methods, \ie{} the lower limit in the precision of our method if one
	takes Monaghan's as fiducial, amounts to 0.01, in agreement with the result
	shown in Figure \ref{fig:comp}.
	\begin{figure}
   		\centering
   			\includegraphics[angle=-90, scale=0.35]{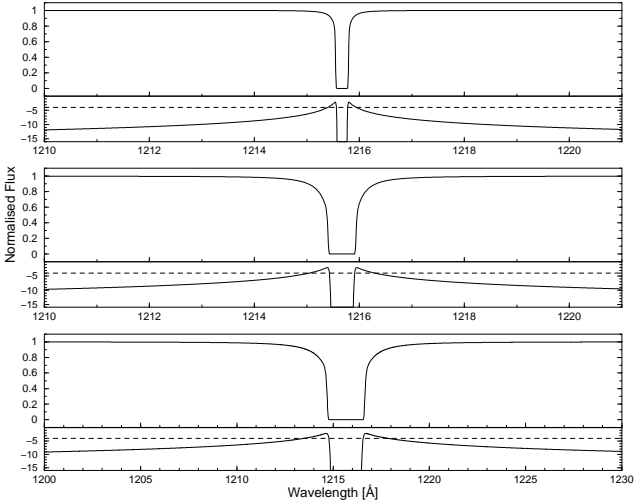}
			\caption[]{\small Examples of \emph{Worst Scenarios}: \lya{}
				absorption line profile synthesised using our method (upper
				panel in each row) and corresponding logarithmic difference with
				respect to our fiducial function (lower panel in each row) for
				the parameter pairs $(\log \nhi,b) = (17.0,10.0)$, $(18.0,20.0)$,
				and $(19.0,70.0)$, from top to bottom, for which the maximum
				discrepancy with respect to the fiducial value of $H$ in Figure
				\ref{fig:map} is largest. The maximum difference of the	order of
				0.01 shown in Figure \ref{fig:map} amounts to an extremely narrow
				wavelength range between Gaussian core and Lorentzian wings. Away
				from these ranges, the accuracy improves significantly. For
				reference, and for comparison with Figure \ref{fig:map}, a line
				corresponding to a constant logarithmic difference of -4 dex has
				been included in each of the lower panels.}
			\label{fig:dla}
	\end{figure}

	As can clearly be seen, the (lower limit in the) precision of our method
	depends both on $b$ and $\nhi$. While the dependence on $b$ extends to the
	whole range $10.0 - 100.0 \, \mathrm{km \, s^{-1}}$, the dependence on $\nhi$
	is limited to the range $\log \nhi = 16 - 20$ dex. For a fixed $b$, there is
	a regime of values around 0.01 with a width of nearly 2 dex, which gives as a
	result a narrow region of low accuracy all across the plane. Along this
	stripe, the difference reaches its highest value of the order of 0.01 or
	smaller for the combinations (low $b$, low $\nhi$) or (high $b$, high
	$\nhi$), in the $b-$ and $\nhi-$ranges stated above. Outside this stripe,
	the accuracy increases dramatically to values of the order	of $10^{-4}$ or
	even better.
	
	The origin of any inaccuracy in our method is obviously the fact that terms
	of order $n \geq 2$ have been neglected in the series (\ref{eq:vhf2}), and
	furthermore, that the second term is this series has been approximated as
	well. In particular, the origin of the 'low-accuracy' stripe on the $(\log
	\nhi,b)$-plane can qualitatively be understood in terms of the functional
	dependence of $\delta V$ on $b$ and $\nhi$. Considering that both our method
	and Monaghan's take the zeroth order term exactly into account, it is
	legitime to state that the quantity $\Delta H \equiv H_{our} - H_{Mon}$ has
	a least a dependence of first order on $a$, \ie{}
	\begin{math}
		\Delta H (a,x) = a \cdot h(a,x)
	\end{math}
	where $h$ is a function that may be of zeroth order in $a$. Hence, using
	equation (\ref{eq:tau}) and this last expression, it follows that
	\begin{displaymath}
		\Delta \tau (\lambda) = C'_i \cdot \frac{\mathrm{N_{HI}}}{b^2} \cdot
		h(a,x) \, , \quad \mathrm{with} \quad C'_i \equiv
		\frac{e^2}{4\,\sqrt{\pi}\,m_{e}\,c} \,\lambda^2_i \, f_i \, \Gamma_i \, .
	\end{displaymath}
	As can be seen, $\tau$ strongly depends on the ratio $\frac{ \mathrm{N_{HI}}
	}{b^2}$. Therefore, an increase in $b$ of one order of magnitude (from 10 to
	100 $\mathrm{km \, s^{-1}}$) is nearly compensated (in the sense that the
	value of $\tau$ remains nearly constant) by an increase in $\nhi$ of two
	orders of magnitude, which accounts for the stripe of 2 dex in column density
	seen on the	$(\log \nhi,b)$-plane. Why this happens precisely between $\log
	\nhi = 16 - 20$ dex, as well as the shape of this stripe, is non-trivially
	related to the exact value of the constant $C'_i$, the fact that $h$ may
	depend also on $b$ through the damping parameter $a$, and the fact that the
	$\Delta V$ depends effectively not on $\Delta \tau$, but on
	$\mathrm{e}^{-\Delta \tau}$.
	
	It is worth mentioning that, for higher-order Lyman transitions, the
	precision of our method to compute Voigt profile has the same behaviour on
	the $(\log \nhi,b)$-plane, and is the same as or even better than the
	precision of the \lya{} transition shown here. The reasons for this are,
	first of all, that the functional form of the absorption coefficient is
	obviously the same, irrespective of the transition. In addition, the lowest
	precision possible of 0.01 is the same for the whole range in $a$ spanned by
	the Lyman transitions, according to Figure \ref{fig:comp}. Furthermore,
	higher transitions have lower damping parameters and our approximation is
	better the lower $a$, as already mentioned. Finally, since the constant
	$C'_i$ is smaller the higher the order of the transition, the critical range
	of lowest precision is shifted to higher column densities and higher Doppler
	parameters. Since the ranges in $\nhi$ and $b$ are fixed for intergalactic
	\hi{}, this has the net effect of increasing the high-precision region on the
	$(\log\nhi,b)$-plane--\ie{} the region to the left of the stripe--for higher
	order transitions. In other words, the precision of our method improves from
	\lya{} to higher Lyman transitions.
	
	Even though a discrepancy of the order of 0.01 when using our method to
	model Voigt profiles may seem large, we want to emphasise again that this is
	merely an \emph{lower} limit for the precision in the \emph{entire}
	wavelength range $\lambda \in [100,2300]$ \AA{}, in the case of \lya{}. It
	turns out that the range in wavelength for which the accuracy is lowest is
	negligible for practical purposes. In order to show this, we first choose
	three points on the $(\log \nhi,b)$-plane along the stripe of lowest
	precision, \ie{} for which the maximum difference is largest. Using these
	parameters, we synthesise \lya{} absorption profiles using our method and
	Monaghan's, and compute again for each of these ''worst scenarios'' the
	quantity $\delta V$ for the whole wavelength range. The result is shown in
	Figure \ref{fig:dla}. Each row corresponds, from top to bottom, to the
	parameter pairs $(\log \nhi,b) = (17.0,10.0)$, $(18.0,20.0)$, and
	$(19.0,70.0)$ chosen along the low-precision stripe. The upper panel of each
	row shows the \lya{} absorption profile synthesised using our method, and the
	corresponding lower panel gives the logarithmic difference $\log_{10} \delta
	V$ between our method and Monaghan's as a function of wavelength. As can be
	seen, the smallest discrepancies are given at the line cores, as expected,
	since in this regime the zeroth order term dominates and both our approach
	and Monaghan's exactly take this term into account. The largest
	discrepancies, of the order of 0.01, are present in an extremely narrow
	range of $\Delta \lambda \approx 0.06$ \AA{} for $(\log \nhi,b) =
	(17.0,10.0)$, of $\Delta \lambda \approx 0.17$ \AA{} for $(\log \nhi,b) =
	(18.0,20.0)$, and of $\Delta \lambda \approx 0.53$ \AA{} for $(\log \nhi,b) =
	(17.0,20.0)$. This discrepancies are found	at the boundaries between
	Gaussian core and Lorentzian wings, due to the fact that our method neglects
	terms of order $n \geq 2$, which dominate the behaviour of $H$ in that
	regime. Note, however, that the difference rapidly drops with increasing
	distance (in \AA{}) from the line center to values of the order of \eg{}
	$10^{-10}$ at a distance $\Delta \lambda \approx 5$ \AA{} for the first two
	rows, and $\Delta \lambda \approx 15$ \AA{} for last row. Hence, the
	effective accuracy of our method is far better than 0.01 in the wavelength
	range shown here. For the same reason mentioned in the last paragraph, the
	precision for higher order Lyman  transitions is of the same order as or even
	better than for the \lya{} transition shown here.

\section{Application} \label{sec:app}

	As a further test of the quality of our approximation to model Voigt
	profiles, and to show its accuracy in a less academic situation as in the
	last section, we consider fitting a synthetic spectrum to a real quasar
	absorption spectrum with a population of intergalactic \hi{}
	absorbers spanning a representative range in $b$ and $\nhi$ along a random
	line-of-sight. For this purpose we use the observed spectrum of the
	quasistellar source QSO J2233-606, a relatively bright ($B = 17.5$) quasar
	at an intermediate redshift $z_{em} = 2.238$.
	
	The spectrum of the source QSO J2233-606, centered at the HDF-S, was
	obtained during the Commissioning of the UVES instrument at the VLT Kueyen
	Telescope and reduced at the Space Telescope European Coordinating Facility.
	The high-resolution spectroscopy (R $\approx$ 45000) was carried out with
	the VLT UV-Visual Echelle Spectrograph (UVES). The data were reduced in the
	ECHELLE/UVES context available in MIDAS. The final combined spectrum has
	constant pixel size of 	0.05 \AA{} and covers the wavelength range
	3050-10000 \AA{}. The S/N ratio of the final spectrum is about 50 per
	resolution element at 4000 \AA{}, 90 at 5000 \AA{}, 80 at 6000 \AA{}, 40
	at 8000 \AA{}. The data used here are publicly available and were retrieved
	from www.stecf.org/hstprogrammes/J22/J22.html, in its version of November
	23, 2005.
	
\begin{figure}
		\centering
			\includegraphics[angle=-90,scale=0.35]{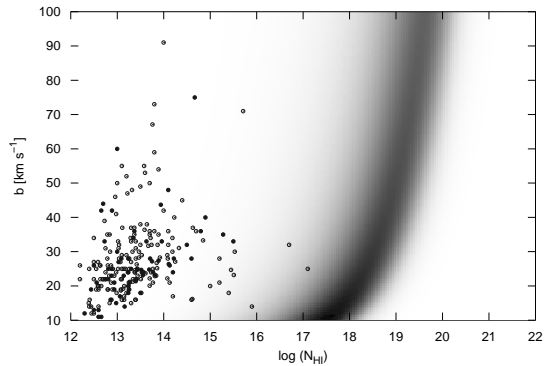}
			\caption{\small Distribution in column density and Doppler parameter
				of the absorbing systems along the line-of-sight towards the
				source QSO J2233-606. This particular LOS contains a total of 264
				absorbing systems with $b \in [10,100] \, \mathrm{km \, s^{-1}}$
				in the range $3050-10000$ \AA{} (open circles), 69 from which are
				in the wavelength range $3340 - 3530$ \AA{} (filled circles). The
				data was taken from \citet{cri}. For reference and comparison,
				the contrast diagram shown in Figure \ref{fig:map} has also been
				included.}
			\label{fig:NHI_b}
	\end{figure}

	With help of the MIDAS package FITLYMAN \citep{fon}, which performs line
	fitting through $\chi^2$ minimization of Voigt profiles, \citet{cri}
	determined the redshifts, column densities, and Doppler widths of the
	identified absorption features imprinted in the spectra of QSO J2233-606. In
	this way, they found that the line of sight to QSO J2233-606 intersects a
	total of 270 \lya{} Forest clouds, and identified other 24 absorption systems
	containing metal lines. The \lya{} absorbers span a range in column density
	of $10^{12.20} - 10^{17.10} \, \mathrm{cm^{-2}}$, and a range in Doppler
	parameters of $1.0 - 111.0 \, \mathrm{km \, s^{-1}}$. The distribution of the
	parameter pairs $(\nhi,b)$ for these systems for $b \in [10,100] \,
	\mathrm{km \, s^{-1}}$  is shown in Figure \ref{fig:NHI_b}.

\begin{figure*}
		\begin{center}
			\vspace{1.5cm}
			\includegraphics[width=0.9\linewidth,angle=-90,scale=0.85]{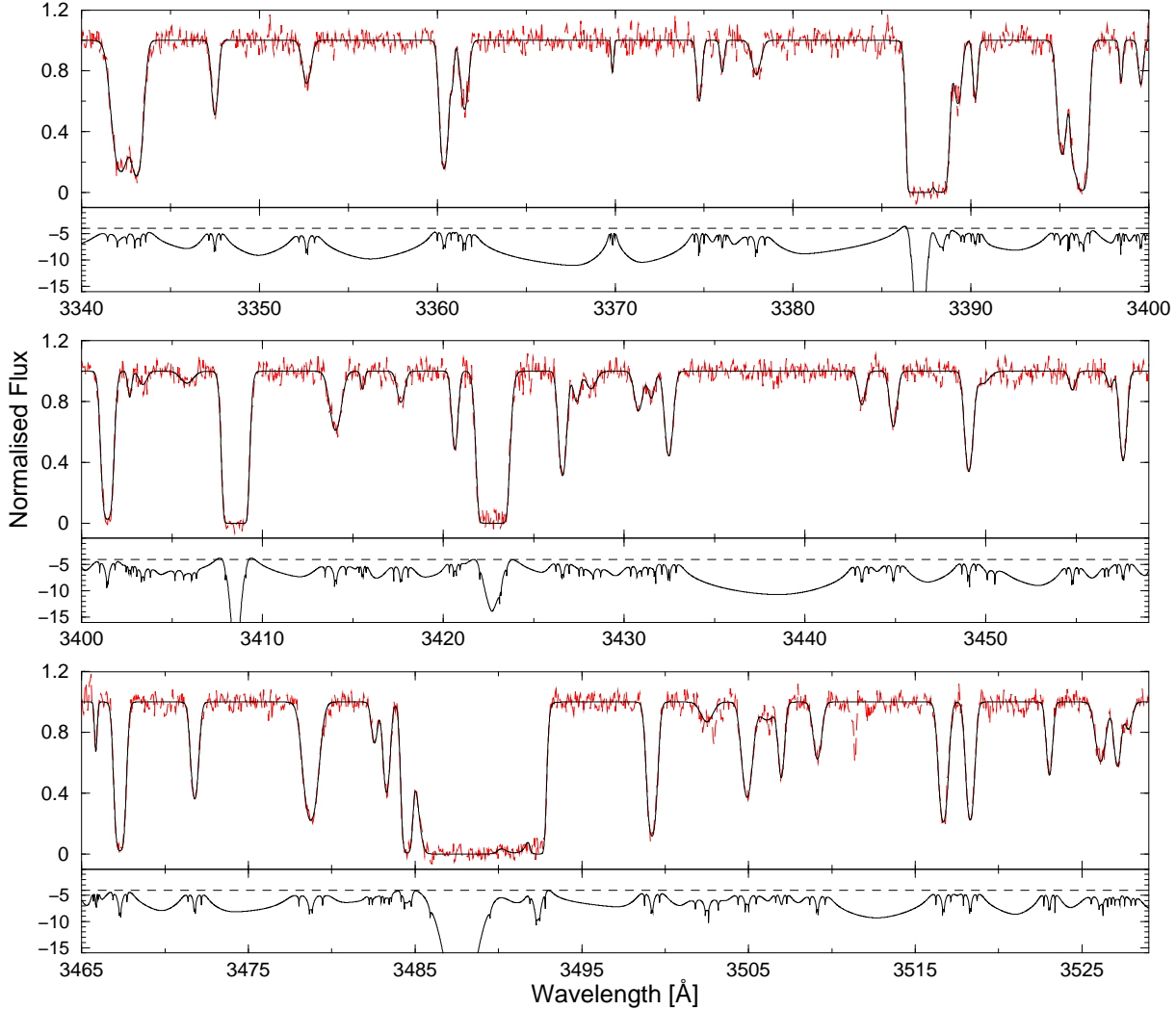}
		\end{center}
		\vspace{1cm}
		\caption{\small Observed (dashed line) and synthetic (solid line)
			spectrum of the quasar HDF-S QSO J2233-606 (upper panel in each row)
			in the wavelength range $[3340,3530]$ \AA{} (cf. text for reference).
			The synthetic spectrum was generated using our approximation to	the
			\VHf{} and the list of line parameters obtained by \citet{cri}. The
			lower panels show the logarithmic difference between our synthetic
			spectrum and one generated using the same list of line parameters and
			Monaghan's algorithm to calculate Voigt profiles. For reference, we
			include in the difference panels an horizontal line corresponding to
			a logarithmic difference of -4 dex.}
		\label{fig:specdiff}
	\end{figure*}

	Using this list of \lya{} line parameters $(z_{abs}, \ b, \ \nhi)$, we
	generate a synthetic spectrum of the QSO J2233-606 in the wavelength range
	$3340 - 3530$ \AA{} with a higher resolution than that of the observed
	spectrum of $\Delta \, \lambda = 0.01$ \AA, using eqs. (\ref{eq:tau}),
	(\ref{eq:h_13}), (\ref{eq:h_14}), and  (\ref{eq:vhf4}). We synthesise a
	second spectrum using Monaghan's algorithm, and compute the logarithmic
	difference $\delta V$ between these two synthetic spectra in the same fashion
	as in the previous section. In this way, we test again our method against the
	highest-precision method available, for a typical range in column densities
	and Doppler parameters present in a QSO spectrum. The result is shown in
	Figure \ref{fig:specdiff}. The upper panels of each row show the observed
	spectrum and the spectrum synthesised using our method, whereas the lower
	panels show the logarithmic difference between both synthetic spectra. We
	choose to cut off the logarithmic difference at $10^{-16}$, since differences
	smaller than these are out of the range of the highest available numerical
	precision. It can be seen again, as in Figure \ref{fig:dla}, that the
	smallest discrepancies are given at the line cores, and the largest, of the
	order of $10^{-4}$, are given at the wings (cf. discussion of Figure
	\ref{fig:dla}, Section \ref{sec:mod}). As can be seen from the column density
	and Doppler parameter distribution in Figure \ref{fig:NHI_b}, the largest
	discrepancies in this wavelength range are consistent with the maximum
	absolute differences shown in Figure \ref{fig:map}. Note that in the spectral
	regions where no apparent absorption features are found, the logarithmic
	difference does \emph{not} fall to $-\infty$, as one would naively expect.
	These features are present in Figure \ref{fig:dla} as well. The reason for
	this 'valley-shaped' features is that, even though having small values away
	from the line center, the function $H$ does \emph{not} fall to zero, and thus
	in these regions the wings of two or more lines overlap. Strictly speaking,
	in these regions there is always some absorption left, \ie{} $\tau < 1$, and
	different methods to compute $H$ will account for this effect differently.
	Again, since our method neglects terms of order $n \geq 2$, the absorption in
	these regimes differs from its fiducial value. The lack of this terms in our
	approximation to $H$ is also pointed out pictorially by the local maxima
	symmetrically placed around the deeps corresponding to the logarithmic
	difference at the line cores.
%
\section{Summary} \label{sec:summ}

	The absorption lines imprinted in the spectra of background sources yield
	a wealth of information about the physical and chemical properties of the
	intervening absorbing material, as is the case of intervening neutral
	hydrogen (\hi{}) systems embedded in the intergalactic medium (IGM) and
	associated with galaxies and larger structures. In order to extract the
	desired information from these absorption lines, their profiles have to be
	modeled in a proper way. In the case of absorption features found on QSO
	spectra, absorption line profiles are best modeled by Voigt profiles, which
	are mathematically given in terms of the \VHf{}.
	
	In this work, we presented a simple analytic approximation to the \VHf{}
	with which Voigt profiles can be modeled for an arbitrary range in
	wavelength (or frequency), column densities up to $10^{22} \
	\mathrm{cm}^{-2}$, and for damping parameters satisfying $a \lesssim
	10^{-4}$. Starting with an exact expression for the \VHf{} in terms of
	Harris' expansion that is valid for $a < 1$, we showed that the zeroth order
	term of this series, a Gaussian function, is suitable for modeling
	absorption line profiles emerging in a medium with low column density
	$\nhi \lesssim 10^{15} \mathrm{cm}^{-2}$. However, for higher column
	densities, terms of higher order have to be taken into account. A key point
	leading to this conclusion is the fact that it is not the damping parameter
	alone, but rather the factor $a \cdot \nhi$ that determines to which extent
	terms of order higher than zeroth in Harris' expansion may or may not be
	neglected. We showed that the departure of the actual \VHf{} from the first
	two terms in Harris' expansion is of the order of $10^{-7}$ or less for
	an arbitrary wavelength range and $a \lesssim 10^{-4}$. Hence, we concluded
	that with an approximation to first order in $a$ to the \VHf{} Voigt profiles
	can be modeled with moderate to high accuracy.
		
	On this basis, we obtained a simple analytic expression for the \VHf{} and
	consequently for the absorption coefficient of intergalactic \hi{}, in terms of
	an approximation for the second term $H_1$ of Harris' expansion.  The main
	advantages of the analytic expression we presented here are, first, that it
	is valid for an arbitrary wavelength range, in the sense that is has no
	singularities. In addition, it is simple and flexible to implement in a
	variety of programming languages to numerically compute Voigt profiles with
	moderate speed and moderate to high accuracy. As a matter of fact, our method
	to compute the \VHf{} is faster with respect to other known standard methods,
	for instance, Huml\'\i $\check{\textnormal{c}}$ek's or Monaghan's algorithm.
	Furthermore, our approximation reaches an accuracy of $10^{-4}$ or better in
	a wide wavelength range, and of the order of than $10^{-2}$ only a negligible
	wavelength interval, for values of $a$ and $\nhi$ characteristic to
	intergalactic \hi{} absorbers. Our method thus offers a great compromise
	between speed, accuracy, and flexibility in its implementation.

	Even though we did not extend our discussion in this work to other
	transitions	typically present in quasar absorption spectra and associated
	to \hi{} absorbers, such as metal lines, our method to synthesise Voigt
	profiles can certainly be applied to most of these elements as well, since
	their column densities are obviously the same, and their ranges in $a$
	strongly overlap with the range of $a$ for intergalactic \hi{}, for	which
	our approximation to the \VHf{} is valid. As a matter of fact, our
	approximation is valid to model absorption Voigt profiles found in any type
	of spectrum (stellar, solar, etc.), which arise in a medium whose damping
	parameter and column density satisfy $a \lesssim 10^{-4}$ and $\nhi \leq
	10^{22} \, \mathrm{cm}^{-2}$, respectively.


\section*{Acknowledgments}	\label{sec:aknow}

	I thank Uta Fritze-v.Alvensleben and the G\"ottingen Galaxy Evolution Group
	for encouraging comments. Special thanks to the referee M.T. Murphy for
	providing his computational routine to compute Voigt profiles, for pointing
	out some additional references, and for very helpful suggestions which helped
	to significantly improved the presentation of our results. This project was
	partially supported by the \textit{Mexican Council for Science and
	Technology} (CONACYT), the G\"ottingen Graduate School of Physics (GGSP), and
	the \textit{Georg-August-University of G\"ottingen}.


\label{lastpage}


\begin{thebibliography}{}

	\bibitem[\protect\citeauthoryear{Abramowitz \& Stegun eds.}{1965}]{hmf}
		Abramowitz, M. \& Stegun, I.A., eds. 1965,
		\textit{Handbook of Mathematical Functions} National Bureau of
		Standards

	\bibitem[\protect\citeauthoryear{Cristiani \& D'Odorico}{2000}]{cri}
		Cristiani, S., \& D'Odorico, V. 2000,
		AJ, 120, 1648 

	\bibitem[\protect\citeauthoryear{Dav\'e \ea}{1997}]{dav}
		Dav\'e, R., Hernquist, L., Weinberg, D.H., \& Katz, N. 1997,
		ApJ, 477, 21

	\bibitem[\protect\citeauthoryear{Dawson}{1898}]{daw}
		Dawson, H.G. 1898,
		Proc. London. Math. Soc., 29, 519

	\bibitem[\protect\citeauthoryear{Finn \& Mugglestone}{1965}]{fin}
		Finn, G.D. \& Mugglestone, D. 1965,
		MNRAS, 129, 221

	\bibitem[\protect\citeauthoryear{Fontana \& Ballester}{1995}]{fon}
		Fontana, A., \& Ballester, P. 1995,
		The ESO Messenger, 80, 37 

	\bibitem[\protect\citeauthoryear{Forster}{1983}]{for} Forster, O. 1983,
		\textit{Analysis}, 4. Auflage, Band 1, Vieweg Verlag
	
	\bibitem[\protect\citeauthoryear{Gregg \ea}{2000}]{gre}
		Gregg, M.D., Wisotzki, L., Becker, R.H., Maza, J., Schechter, P.L.,
		White, R.L., Brotherton, M.S., \& Winn, J.N. 2000,
		AJ, 119, 2535		 
	
	\bibitem[\protect\citeauthoryear{Harris}{1948}]{har}
		Harris, D.L. III 1948,
		ApJ, 108, 112
		
	\bibitem[\protect\citeauthoryear{Hjerting}{1938}]{hje}
		Hjerting, F. 1938,
		ApJ, 88, 508

	\bibitem[\protect\citeauthoryear{Hu \ea{}}{1995}]{hue}
		Hu, E.M., Kim,  T.-S., Cowie, L.L., \& Songaila, A. 1995,
		AJ, 110, 1526


	\bibitem[\protect\citeauthoryear{Huml\'\i
		$\check{\textnormal{c}}$ek}{1982}]{hum}
		Huml\'\i $\check{\textnormal{c}}$ek, J. 1982,
		JQSRT, 27, 437
	
	\bibitem[\protect\citeauthoryear{Kielkopf}{1973}]{kie}
		Kielkopf, J.F. 1973,
		JOSA, 63, 987
	
	\bibitem[\protect\citeauthoryear{Kim \ea{}}{1997}]{kim97}
		Kim,  T.-S., Hu, E.M., Cowie, L.L., \& Songaila, A. 1997,
		AJ, 114, 1
	
	\bibitem[\protect\citeauthoryear{Kim \ea{}}{2001}]{kim01}
		Kim,  T.-S., Cristiani, S. \& D'Odorico, S. 2001,
		A \& A, 373, 757
	
	\bibitem[\protect\citeauthoryear{Kim \ea{}}{2002}]{kim02}
		Kim,  T.-S., Carswell, R.F., Cristiani, S., D'Odorico, S. \& Giallongo,
		E. 2002,
		MNRAS, 335, 555
	
	\bibitem[\protect\citeauthoryear{Mihalas}{1970}]{mih}
		Mihalas, D. 1970,
		Stellar Atmospheres,
		W.H. Freeman and Company, San Francisco

	\bibitem[\protect\citeauthoryear{Miller \& Gordon}{1931}]{mil}
		Miller, W.L. \& Gordon, A.R. 1931,
		J.Phys.Chem, 35, 2785

	\bibitem[\protect\citeauthoryear{Monaghan}{1971}]{mon}
		Monaghan, J.J., 1971,
		MNRAS, 152, 509

	\bibitem[\protect\citeauthoryear{Morton}{2003}]{mor}
		Morton, D.C., 2003
		ApJ Suppl. 149, 205
	\bibitem[\protect\citeauthoryear{Murphy}{2002}]{mur}
		Murphy, M.T., 2002,
		\emph{Probing Variations in the Fundamental Constants With Quasar
		Absorption Lines},
		PhD thesis, University of New South Wales

	\bibitem[\protect\citeauthoryear{Rao}{2005}]{rao}
		Rao, S.M. 2005,
		\textit{Probing Galaxies Through Quasar Absorption Lines},
		Proceedings IAU Colloquium No. 199,
		Williams, P.R., Shu, C., M\'enard, B. eds. 
		
	\bibitem[\protect\citeauthoryear{Rauch}{1998}]{rau}
		Rauch, M. 1998,
		ARA\&A, 36, 267
	
	\bibitem[\protect\citeauthoryear{Richter \ea{}}{2005}]{ric}
		Richter, P., Fang, T., \& Bryan. G.L. 2005,
		A \& A, astro-ph/0511609

	\bibitem[\protect\citeauthoryear{Spitzer}{1978}]{spi}
		Spitzer, L. 1978,
		Physical Processes in the Interstellar Medium, 
		Wiley: New York, p. 143
			
	\bibitem[\protect\citeauthoryear{Walshaw}{1955}]{wal}
		Walshaw, C. D. 1955,
		Proc. Phys. Soc., A68, 530
		
	\bibitem[\protect\citeauthoryear{Whiting}{1968}]{whi}
		Whiting, E.E. 1968,
		JQSRT, 8, 1379

	\bibitem[\protect\citeauthoryear{Zhang \ea{}}{1997}]{zha}
		Zhang, Y., Anninos, P., Norman, M.L., \& Meiksin A. 1997,
		ApJ, 485, 496

\end{thebibliography}
\end{document}